\documentstyle[11pt,newpasp,twoside,epsf]{article}
\def\edcomment#1{\iffalse\marginpar{\raggedright\sl#1\/}\else\relax\fi}
\reversemarginpar

\begin{document}

\title{The Hanle Effect as a Magnetic Diagnostic}

 \author{Richard Ignace}

\affil{University of Wisconsin, Department of Astronomy, 5534
Sterling Hall, Madison, WI 53706, USA}

%\author{Ima Co-Author}
%\affil{The Name of My Institution, The Full Address of My Institution}

\begin{abstract}
The physics of the Hanle effect is briefly reviewed, and its application
as a diagnostic of hot star magnetic fields is described.  Emphasis is
given to the practicalities of using spectropolarimetry of resolved wind
emission lines to infer information about the circumstellar magnetic
field strength and its geometry.  A model for a weakly magnetized wind
from ``WCFields'' theory is used as the backdrop for investigating
polarized line profile effects for P~Cygni resonance lines using
a kind of ``last scattering approximation''.  Model results are
presented for a typical P~Cygni line that forms in a spherical wind.
Significant line polarizations of a few tenths of a percent can result
for circumstellar fields of order 100~G.  Information about the field
topology and surface field strength is gleaned from the Stokes-$Q$ and
$U$-profiles.  Simplistically, the $Q$-profile polarization is governed
by the field strength, and the $U$-profile symmetry (whether symmetric
or anti-symmetric or even zero) is governed by the field geometry.
\end{abstract}

\section{Introduction}

The Hanle effect is a magnetic field diagnostic that is still relatively
unknown among stellar astronomers.  The Effect refers to the modification
(either an increase or decrease, as well as a change in position angle)
of resonance line scattering polarization in the presence of a magnetic
field, which begins to have an influence at fairly modest (even small)
field strengths of just a few Gauss up to around 300~Gauss.  Experiments
to describe the polarization from resonance line scattering date back
primarily to the early 1900's (e.g., see Mitchell \& Zemansky 1934).
The influence of a magnetic field on the line polarization was explained
first by a young physicist named Hanle for his dissertation (Hanle 1924).
He described the effect of the magnetic field in semi-classical terms
as arising from the precession of an atomic damped harmonic oscillator.
Quantum mechanically, the Effect is understood in terms of an interference
effect that occurs when the degeneracy of the magnetic sublevels in the
excited level is partially lifted, and has applications to many topics
in atomic physics (e.g., see the review by Moruzzi \& Strumia 1991).

Astrophysically, the Hanle effect has been applied almost exclusively
in the solar context, especially to measuring magnetic fields in the
solar chromosphere and corona, and particularly in prominences and filaments
(e.g., Lin, Penn, \& Kuhn 1998).  A detailed description of the physics
of the Hanle effect and polarized radiation transport with application
to solar studies appears in Stenflo (1994).  I began investigating
the Hanle effect as a means for measuring circumstellar
magnetic fields, especially in hot stars, as part of my graduate thesis
(1996).  This early work appears in Ignace, Nordsieck, \& Cassinelli
(1997), Ignace, Cassinelli, \& Nordsieck (1999), and Ignace
(2001).  Development of the Hanle diagnostic continues, and here I wish
to summarize its status, as well as observational prospects.  First,
I want to begin with a consideration of the more familiar Zeeman
effect to use as a springboard for discussing the Hanle effect.

\section{The Zeeman Effect for Detecting Circumstellar Magnetic Fields}

The strength of the Zeeman effect and associated circularly polarized
emission depends on the Zeeman splitting $\Delta \lambda_B$,
as given by

\begin{equation}
\Delta \lambda_B = 4.67\times 10^{-9} \mu {\rm m} \; \cdot \; g_{\rm L}\,
        \lambda^2\, B ,
        \label{eq:zeeman}
\end{equation}

\noindent with $\lambda$ in microns, $B$ in Gauss, and $g_{\rm L}$ the
effective L\'{a}nde factor to generalize the expression.  A line in the
visible band and a modest magnetic field of around 100~G yields a Zeeman
splitting of a mere 75 m s$^{-1}$.  This is considerably smaller than
the thermal broadening typical of stellar envelopes, and much smaller
than the velocity broadening in the winds, and this presents a serious
difficulty: the circular polarizations of the two Zeeman split components
are incoherent and oppositely signed, so that their broadening-induced
overlap leads to severe polarimetric {\it cancellation}.  What survives
is a much weaker antisymmetric signal.

\begin{figure}
\plottwo{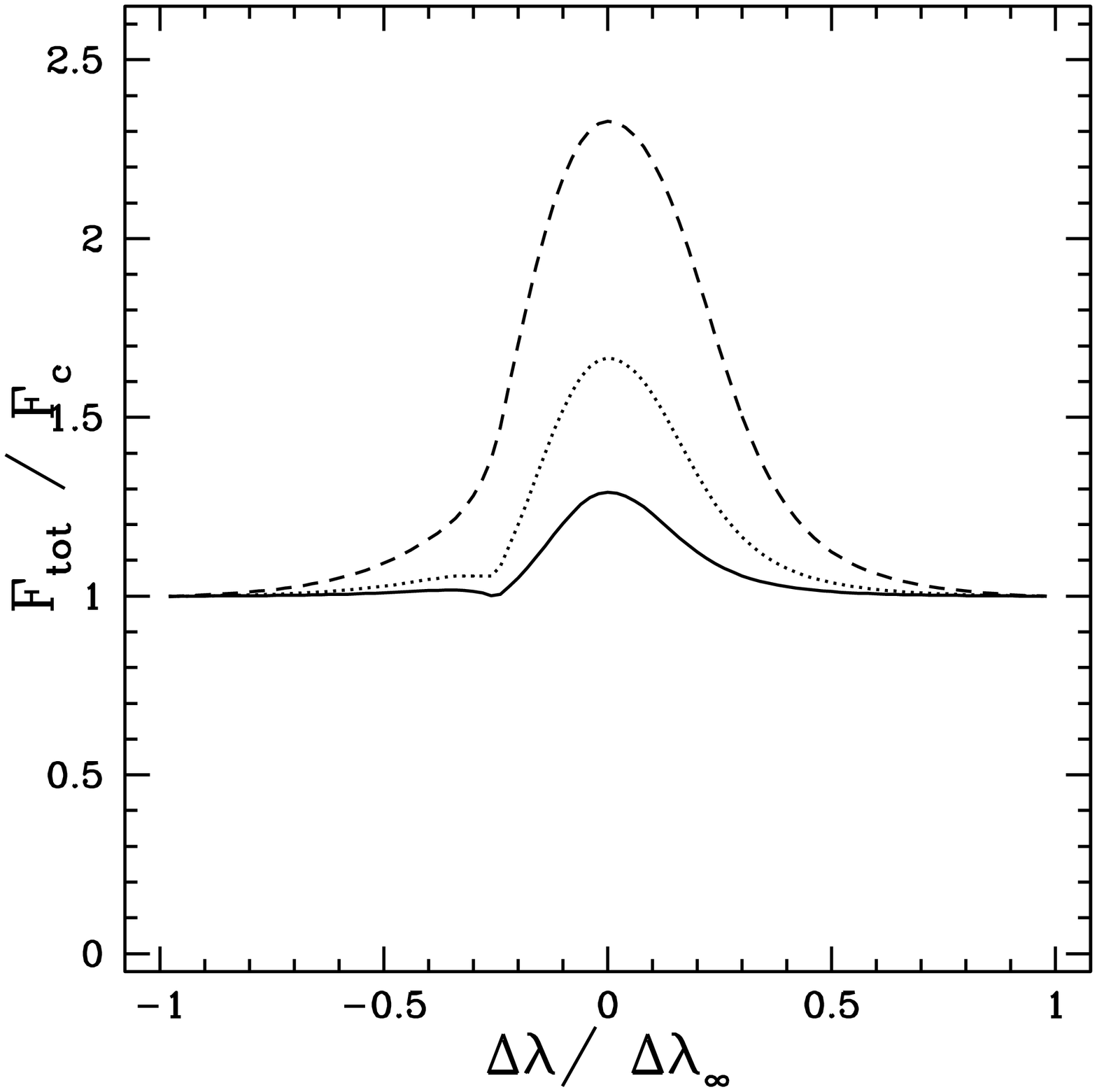}{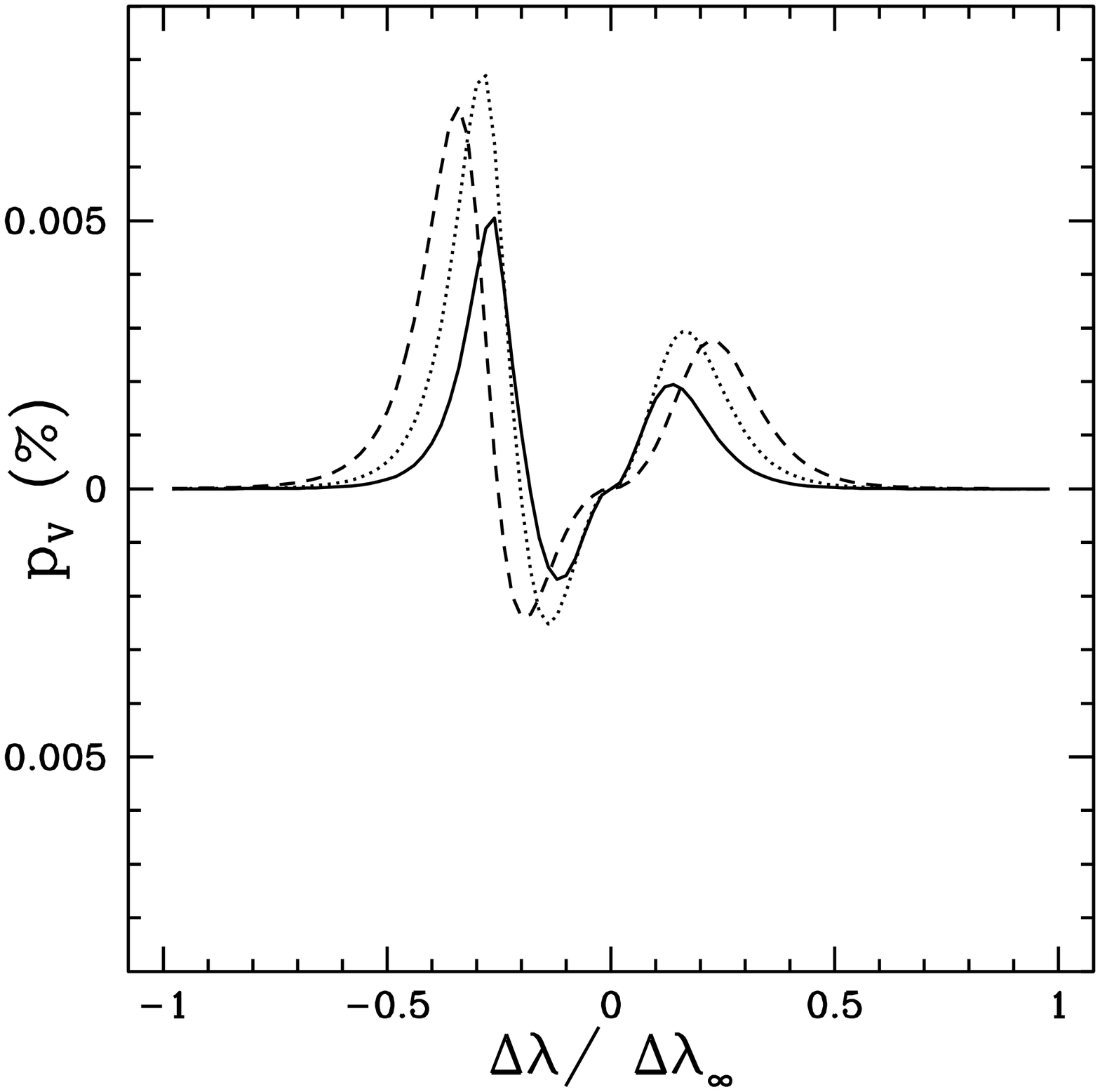}
\caption{At left are continuum normalized recombination lines, each with a
different line optical depth parameter, ranging from moderately thin to
fairly thick.  The magnetic field is assumed to be a split monopole
threading a spherical wind in linear expansion ($v(r) \propto r$).  At right
is the resulting Stokes-$V$ profile shown in percent polarization.}
\label{fig1}
\end{figure}

Following Stenflo (1994), and considering the fields to be ``weak''
(such that the Zeeman splitting is much less than the thermal broadening),
the first order effect is the longitudinal Zeeman effect with Stokes $V$
given by

\begin{equation}
V = -  \Delta \lambda_B\,\cos \gamma\,\left(\frac{dI_0}{d\lambda}
        \right)_{\Delta \lambda}
        \label{eq:V}
\end{equation}

\noindent where $I_0$ is the intensity profile shape in the absence of a
magnetic field and $\cos \gamma = \hat{B} \cdot \hat{z}$, for $\hat{B}$
the magnetic field unit vector and $\hat{z}$ a unit vector directed toward
the observer.  This expression indicates that the circular polarization
will depend on the amount of the Zeeman shift, the orientation of the
observer with respect to the vector magnetic field, and a gradient of
the line profile function.

In the case of a stellar wind, standard expressions for Sobolev theory
are adopted to describe the line profile formation (e.g., Mihalas 1978).
An important concept in this theory is the ``isovelocity zone''.  All of
the wind emission that appears at a given frequency in the line profile
arises from a zone of fixed Doppler shift with respect to the observer.
For example in a radially expanding wind, emission that appears at line
center must arise from a surface coincident with the plane of the sky
and passing through the center of the star, since the flow in that region
has no velocity component along the line-of-sight.

The flux of circular polarization arising from the longitudinal Zeeman
effect at a point in an isovelocity zone will depend on the line-of-sight
gradient of the line source function and the line optical depth (see
Ignace \& Gayley 2003).  The total polarized flux at a given line
frequency then requires an integration over the entire isovelocity zone.
Figure~1 shows an example for mildly thick recombination lines.
The abscissa for each plot is the wavelength shift $\Delta \lambda$
from line center as normalized by the maximum shift for the terminal
speed flow $\Delta \lambda_\infty = \lambda_0 v_\infty /c$.  At left
is the continuum normalized flux, the three lines being for different
optical depths.  At right is the Stokes-$V$ polarization with $p_V =
F_V/F_{tot}$.  The percent polarizations are {\em extremely} small, at
less than 0.01\%!  These low values result primarily from the fact that
the gradient factor introduces a scaling that is inversely proportional
to the line broadening, which in this case is $\Delta \lambda_B / \Delta
\lambda_\infty \ll 1$.  In contrast, the Hanle effect is a different
method for measuring magnetic fields that is especially well-suited for
the circumstellar case.

\begin{figure}[h!]
\epsfxsize = .6\hsize
\hspace{1in}\epsfbox{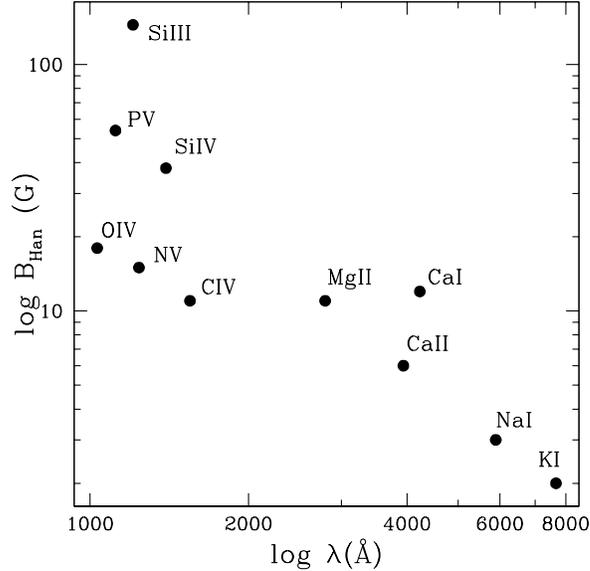}
\caption{Plotted is the logarithm of the Hanle field sensitivity $B_{Han}$
against line wavelength.  The Hanle field is defined such that $2g_L\omega_L /
A = B/B_{Han} = 1$.  The ion yielding each
resonance line is indicated.  The Ca{\sc i} and Si{\sc iii} lines are
singlets; the rest are Li-like doublets.  Low ionization optical lines
are sensitive to fields in the 1-10~G range, whereas high ionization UV
lines are sensitive to stronger fields in the 10-100~G range.}
\label{fig2}
\end{figure}

\section{Physics of the Hanle Effect}

It should be stressed again that the Hanle effect is a {\em modification}
of the linear polarization that arises from resonance line scattering.
Semi-classically, line scattering can be split into two parts: an
isotropic scattering component and a dipole scattering component,
like free electrons (e.g., Hamiton 1947; Chandrasekhar 1961).
The fraction of the scattering that is dipole-like is referred to as
the ``polarizability'', $E_1$; the isotropic scattering contribution is
given fractionally as $(1-E_1)$.  

One can think of resonance scattering in terms of a damped harmonic
oscillator.  Suppose that a beam of polarized intensity (100\%) impinges
on a sample of scatterers.  The oscillators are excited and oscillate
parallel to the sense of the beam polarization, scattering light 
anisotropically (with peak intensity perpendicular to the direction
of oscillation and zero intensity along that direction).

The introduction of a magnetic field induces a precession of the
oscillators.  There are two important points here.  (a)~The atomic
oscillators no longer ``bob'' in a fixed direction.  Instead, they
begin to oscillate in directions perpendicular to the polarization of
the incident light.  Consequently, one can think of the Hanle effect
as redistributing energy between orthogonal axes of atomic oscillation.
(b)~The oscillators are damped, emitting their absorbed radiation at a
rate given by the Einstein $A$-value for the transition of interest.
Magnetic precession occurs at a rate given by the Larmor frequency,
$\omega_L = e B /m_e c$.  Hence there will be a Hanle effect only if
these two rates are similar.  I define the ``Hanle ratio'' as

\begin{equation}
\frac{g_L \omega_L}{A} = g_L\,\frac{(B/10~G)}{(A/10^8~s^{-1})}.
\end{equation}

\noindent Typical strong lines are sensitive to the Hanle effect for
field strengths in the 1-300~G range (see Fig.~2).  Note also that
different lines will have different Hanle sensitivities.  Finally, if
the Hanle ratio is small, the scattering is essentially non-magnetic
(i.e., little precession occurs over the lifetime of the level), but
if the ratio is large, the Hanle effect is said to be ``saturated''.
Saturated means that the precession is so rapid that little damping
of the oscillators occurs after one full rotation, so that to increase
the field will not alter the Hanle effect.  Diagnostically, one loses
sensitivity to the field strength, but the scattering polarization still
depends on the vector orientation of the magnetic field.  Of course,
if the field becomes very large indeed, the scattering physics shifts
from the Hanle effect to the Zeeman effect.

\section{Application of the Hanle Effect to P Cygni Lines}

There are two main differences that distinguish the use of the Hanle
effect in studies of the Sun versus hot stars.  The first is the fact
that the Sun is a resolved source, and stars are not.  Cassinelli,
Nordsieck, \& Ignace (2001) have also pointed out that in the hot star
wind case, the lines are strongly NLTE and relatively simple to treat,
and the scattering is of a ``flat'' stellar continuum, whereas neither of these
is the case for most applications of the Hanle effect in the Sun.
So to compute polarized profiles of {\it wind emission lines}, standard
Sobolev theory is once again adopted.  For resonance scattering lines,
Sobolev theory yields P~Cygni line profiles that are commonly observed
from stellar winds (see Lamers \& Cassinelli 1999).

\subsection{The Single Scattering Approximation}

The single scattering approximation is simply a physically motivated
convenience whereby the line emission from optically thick portions of
the wind is assumed to be completely unpolarized, and the emission from
optically thin portions is assumed to be described by the source functions
given in Ignace (2001).  The multiple scattering in thick regions is expected
to be depolarized, whereas multiple scattering does not occur in
thin regions.  This approach is closely related to the ``last scattering
approximation'' that is sometimes employed for static media.

\subsection{WCFields Primer}

A realistic model for a magnetized wind is required to produce synthetic
polarized line profiles.  WCFields is an axisymmetric semi-analytic model
for just such a wind that is based on ``wind compression'' (WC) theory.
In WC~theory (Bjorkman \& Cassinelli 1993; Ignace, Cassinelli, \& Bjorkman
1996), the wind flow from a rotating star is described kinematically.
Ignoring pressure gradient terms, and assuming radial wind-driving
forces, one can derive an expression for the streamline flow.  The general
properties of WC theory have been confirmed with hydrodynamic simulations
by Owocki, Cranmer, \& Blondin (1994), although the effects appear to be
inhibited when non-radial accelerations that arise in the line-driving
of hot star winds are included (Owocki, Cranmer, \& Gayley 1996).

However, retaining the basic elements of WC~theory,
Ignace, Cassinelli, \& Bjorkman (1998) extended the method to allow for ``weak''
magnetic fields.  The key assumptions are that the magnetic field
is frozen-in and dominated by the hydrodynamic flow.
Consequently, the known flow geometry determines the magnetic field
topology.  Referring to Ignace et~al.\ (1998) for the derivation, the
expression that determines the vector magnetic field throughout the
flow is

\begin{equation}
{\bf B} = B_*\,\left(\frac{R_*^2}{r^2}\right)\,\left(\frac{d\mu}{d\mu_0}
        \right)^{-1}\,\frac{{\bf V}}{v_{\rm r}},
\end{equation}

\noindent where $B_*$ is the surface field strength, $d\mu/d\mu_0$
is the ``compression factor'' that describes how neighboring streamlines
evolve throughout the wind flow, ${\bf V}$ is the vector velocity in
the co-rotating frame, and $v_{\rm r}$ is the radial velocity.
All of the velocity components and the compression factor are knowns.
The scalar strength of the field is given by

\begin{equation}
B = B_*\,\left(\frac{R_*^2}{r^2}\right)\,\left(\frac{d\mu}{d\mu_0}
        \right)^{-1}\,\sqrt{1+\frac{v_{\rm rot}^2}{v_{\rm r}^2}\,
        \left(\frac{R_*}{r}\sin\vartheta_0-\frac{r}{R_*}\sin\vartheta\right)^2},\end{equation}

\noindent where $v_{\rm rot}$ is the equatorial rotation speed of the
star.  The zero subscript indicates a value at the base of the wind.
The latitude of a streamline varies with radius as flow migrates toward
the equator from both the upper and lower hemispheres.  The field is taken
to be radial at the wind base, but obtains a toroidal configuration at
large radius, with $B\propto r^{-1}\,\sin\vartheta$, like the magnetic
field that is dragged out in the solar wind.  For the slow rotations that
will be considered here, the factor $d\mu/d\mu_0$ varies little from pole
to equator and so is ignored.  Additionally, $B_\vartheta$ is generally
smaller than either $B_{\rm r}$ or $B_\varphi$, and is also ignored.
The resulting model is a spherically symmetric wind with a complex radial and
toroidal magnetic field morphology.

\subsection{Model Line Profile Results}

Using WCFields theory for slowly rotating stars, polarized emission
profiles from P~Cygni lines have been computed and are shown in
Figure~3.  The rotation was chosen to be 8\% of the terminal
speed.  Since most O~star winds have terminal speeds of around 2000
km s$^{-1}$, the equatorial rotation speed would be around 150 km
s$^{-1}$, which is not atypical of observed $v \sin i$ values (Penny
1996).  Especially important is the fact that the wind is assumed
to be spherically symmetric.  In WCFields theory, this model would
have a mild equator-to-pole density contrast of 1.25 asymptotically.
The wind is taken to be explicitly spherical to isolate the influence of
the Hanle effect, since an unresolved spherically symmetric distribution
of scatterers would otherwise yield no net polarization.  The different
profiles are for different surface field strengths as indicated.  The
upper frames are for the Stokes $Q$ profile, and the lower for Stokes~$U$.

The left panels are for the wrong field topology and the right is for
the correct one.  The wrong case assumes a surface magnetic field that
is initially outward radial -- a monopole, which violates the condition
$\nabla$$\cdot {\bf B} = 0$ over a closed surface.  The panels at right
are for a surface field that is initially a split monopole, with field
lines outward radial in the upper hemisphere and inward radial in the
lower hemisphere.  In every case the star is viewed as edge-on.

The two cases are shown to illustrate how the Hanle effect can be used to
glean information about the magnetic field.  What is immediately striking
is that the surface monopole case yields a net $U$-profile, whereas the
surface split monopole case does not.  So the $U$-profile is a sensitive
probe of the field symmetry (such as top-bottom and left-right).

The $Q$-profile also depends on the field symmetry, but simplistically,
the $Q$-profile shows stronger polarizations for larger surface fields,
and can thus be considered a gauge of the surface field strength.
Note that the surface fields rise up to 1500~G in these examples.
Such large values are beyond the Hanle effect and firmly in the Zeeman
regime (Stenflo 1998).  However, in these models, the inner wind is quite
optically thick in the line opacity, and only the outer wind is thin
(e.g., some portions of the wind start to become thin around $2R_*$).
It is only these thin portions that contribute to the Hanle effect in
the single scattering approximation, and at these radii the field is
considerably reduced relative to the surface value.

\begin{figure}
\plottwo{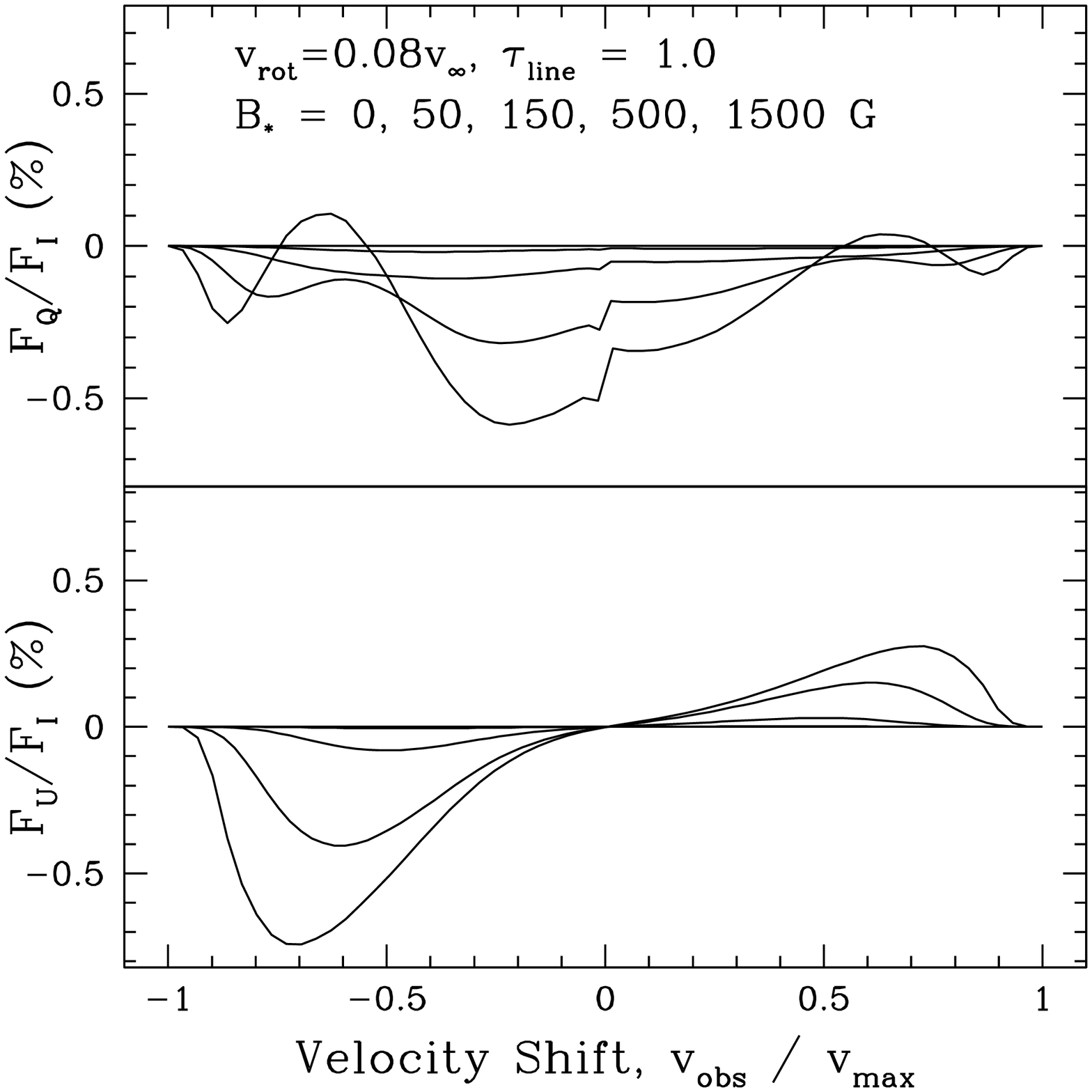}{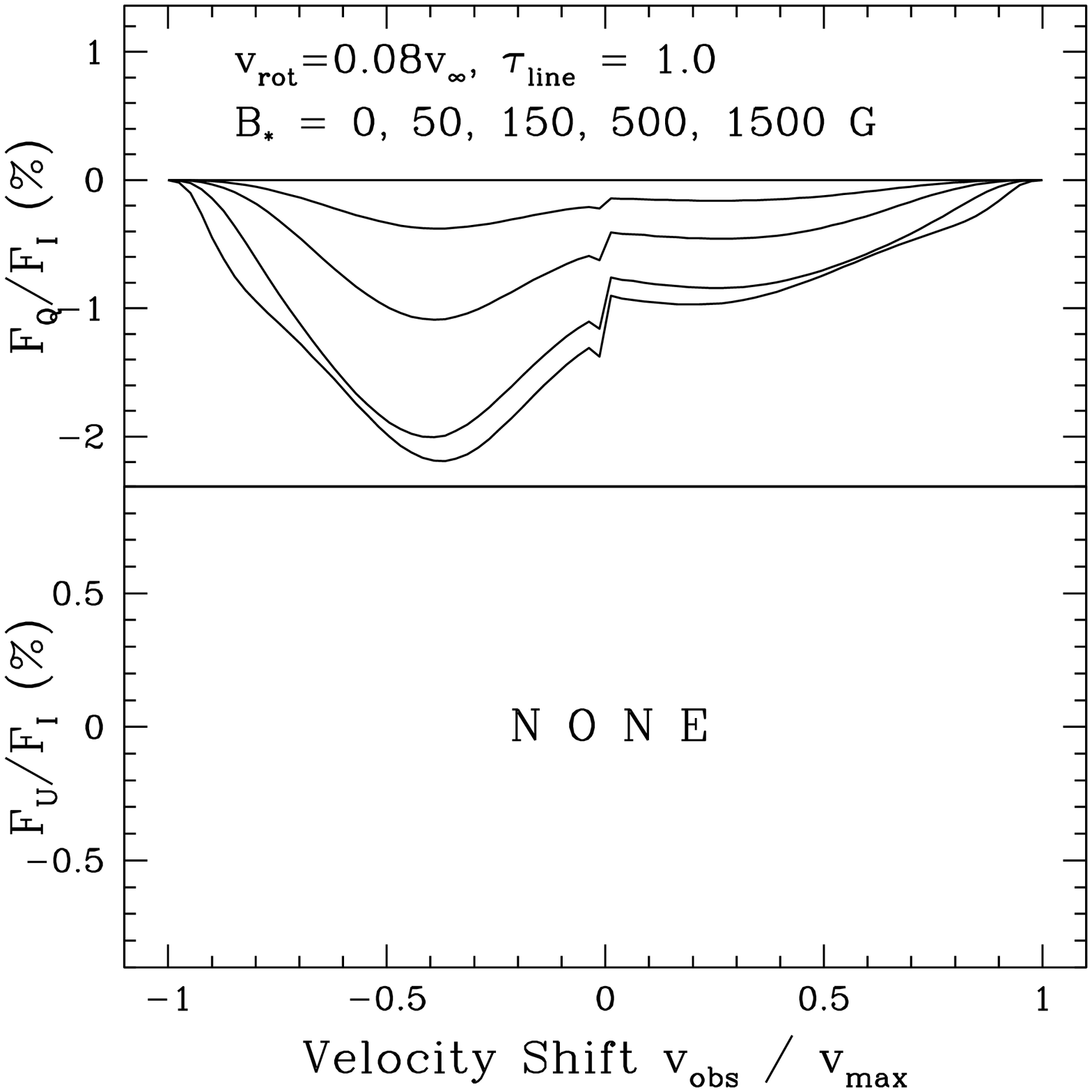}
\caption{Shown are Stokes $Q$ and $U$ polarized line profiles for
{\em spherical} winds and WCFields magnetic geometries.  The stellar rotation,
which determines the field topology, is fixed at 8\% of the terminal
speed, and the star is viewed from an equator-on perspective.  Each line
is for a different surface field strength as indicated, with higher fields
producing stronger line polarizations  The only difference between the
left and right is that in former, the surface field is initially a monopole,
and in the latter it is a split monopole.  Consequently, the $U$-profile
is quite senstive to the symmetry of the field topology, whereas the 
$Q$-polarization generally grows with surface field strength.  }
\label{fig3}
\end{figure}

In summary, significant linear polarizations in the line of a few tenths
of a percent can result from the Hanle effect, and the influence of the
line optical depth is to ensure that for typical P~Cygni lines, the Hanle
effect will probe circumstellar magnetic fields at intermediate radii
of a few $R_*$ in the wind, even if the wind is spherical.  For cases
when the wind is distorted from spherical, as one would expect from a
magnetized flow, polarizations from more than one line, each having a
different Hanle sensivitity, will generally be needed to disentangle
the effects of the magnetic field from that of the non-spherical geometry.

\begin{table}
%\begin{center}
\caption{Some FUSP Targets}
\begin{tabular}{llll}
\tableline
Star & V & $m_{133}^a$ & Comment \\
\tableline
$\zeta$ Ori  & 2.1 & $-$2.5 & {\bf Target for Hanle effect} \\
$\gamma$ Cas & 2.5 & $-$1.6 & Be star \\
$\zeta$ Tau  & 3.0 & $-$0.3 & Be star  \\
$\xi$ Per    & 4.0 & $\;\;$1.0 & O7e rapid rotator \\
HD 93521     & 7.1 & $\;\;$2.8 & O9e rapid rotator \\
EZ CMa       & 6.9 & $\;\;$2.5 & Wolf-Rayet star \\
$\beta$ Ori  & 0.1 & $-$0.8 & Supergiant star \\
$\beta$ Lyr  & 3.5 & $\;\;$2.0 & Binary system \\
\tableline
\tableline
\end{tabular}

%\centerline{$^a$ Apparent magnitude at 133 nm}

\hspace{4em}$^a$ Apparent magnitude at 133 nm
%\end{center}
\end{table}

\section{Observational Prospects}

The Far Ultraviolet SpectroPolarimeter (FUSP) is a sounding rocket payload
that is expected to launch in autumn 2003.  The goal of the mission
will be to make spectropolarimetric measurements in the wavelength
range of 1050--1500~\AA\ for several different targets (see Tab.~1).
With a 50~cm primary, the instrument will have a spectral resolution of
0.65~\AA, corresponding to $\lambda/\Delta \lambda = 1800$ at a wavelength
of 1170~\AA.  This in turn corresponds to a velocity resolution of about
170~km~s$^{-1}$, which is sufficient to resolve wind-broadened lines
with typical full widths of a few thousand km~s$^{-1}$.  Targeting the
star $\zeta$~Ori, which was observed in the FUV with Copernicus, it
is anticipated that FUSP will produce the first detection of the Hanle
effect in a star other than the Sun.

\section{Concluding Remarks}

This contribution has focussed on the Hanle effect in resonance
scattering lines common to hot star winds.  The study of the Effect and
the development of radiative transport techniques is ongoing, especially
in terms of relaxing the single scattering approximation for the line
polarization.  The description presented here ignores the influence
of nuclear spin for the polarizability $E_1$ of a particular transition,
but nuclear spin can have a significant effect on its value (e.g.,
Nordsieck 2001).  We have also ignored the influence of fluorescent
alignment, which affects the populations of the magnetic sublevels.
Fluorescent alignment is also subject to magnetic ``re-alignment''
effects, which are
related to the Hanle effect and operate at much weaker field strengths
(see Nordsieck 2001).  We are investigating the regime in which the
Hanle effect and the magnetic re-alignment physics both apply.

\acknowledgements I want to thank Joe Cassinelli and Ken Nordsieck
for helpful suggestions in preparing this presentation.  Support for
this research comes from a grant from the National Science Foundation
(AST-0098597).

\section*{Discussion}

{\small 

\noindent {\it Wade:}  The Hanle effect is certainly
observed in optical lines in the solar spectrum.  Why do you need to go
to the FUV for hot stars? \\

\noindent {\it Ignace:}  The answer comes down to suitable lines for
suitable stars.  For hot stars, most of the flux is in the UV, and
spectropolarimetry requires lots of photons.  Moreover, the Einstein
$A$-values that set the scale for the magnetic field sensitivity are
greater for shorter wavelength lines, so that UV lines are sensitive to
$\sim 10-300$~G fields, but optical lines to $\sim 1-10$~G.  Finally,
strong resonance scattering lines are found only in the UV for hot stars
(e.g., N{\sc v}, C{\sc iv}, etc), whereas the optical spectrum will have
photospheric absorption or recombination emission lines.  Optical lines
like Na{\sc i} and Ca{\sc ii} can show a Hanle effect, but these low
ions do not exist in hot stars.
}

\end{document}